\documentclass[11pt,twoside]{article}

\usepackage{asp2004}
\usepackage{epsf}
\usepackage{lscape}
\usepackage{graphicx}
\usepackage{subfigure}

\begin{document}
\title{A Progress Report on the Empirical Determination of the ZZ Ceti Instability Strip}   
\author{A. Gianninas, P. Bergeron, and G. Fontaine}   
\affil{D\'epartement de Physique, Universit\'e de Montr\'eal, C.P. 6128, Succ. 
Centre-Ville, Montr\'eal, Qu\'ebec, Canada, H3C 3J7}    

\begin{abstract} 
Although the Sloan Digital Sky Survey has permitted the discovery of
an increasing number of new ZZ Ceti stars, recent published analyses
have shown that there are still many relatively bright (V $<$ 17) ZZ
Ceti stars waiting to be found. We will discuss the discovery of
several such objects in addition to a number of DA stars in which we
detected no photometric variations. These were uncovered as part of an
ongoing spectroscopic survey of DA white dwarfs from the McCook \&
Sion Catalog. By determining the atmospheric parameters of a large
sample of DA stars, we were able to identify objects placed within or
near the empirical boundaries of the ZZ Ceti instability strip. By
establishing the photometric status of these stars, we can use them in
an effort to conclusively pin down the empirical boundaries of the ZZ
Ceti instability strip.
\end{abstract}



\section{Ongoing Survey of the McCook \& Sion Catalog}

Since the early 1990s, our group in Montreal has been carrying out a
systematic study aimed at defining the empirical boundaries of the ZZ
Ceti instability strip. We use quantitative time-averaged optical
spectroscopy to pin down the locations of candidate stars in the
$T_{\rm eff}$-$\log g$ plane, and we follow up on these candidates
with ``white light'' fast photometry to determine whether a target
pulsates or not. The determination of these boundaries is essential if
we are to understand the ZZ Ceti phenomenon as a whole. The question
of the purity of the instability strip is also an important issue for
several reasons. First, a pure strip implies that ZZ Ceti stars
represent an evolutionary phase through which all DA white dwarfs must
pass. It is then possible to apply what is learned from
asteroseismological studies of these stars to the entire class of DA
white dwarfs. Second, knowing the strip is pure allows us to predict
the variability of stars that lie within its boundaries.

In the last few years, we have undertaken a spectroscopic survey of a
subsample of DA white dwarfs from the Catalog of Spectroscopically
Identified White Dwarfs of \citet{mccook99}. We thus obtain optical
spectra for each star and by fitting the Balmer line profiles using a
grid of synthetic spectra generated from detailed model atmospheres we
can accurately measure $T_{\rm eff}$ and $\log g$ for each star. It is
then possible to identify white dwarfs whose atmospheric parameters
place them near or within the ZZ Ceti instability strip. We will
discuss here the latest results of this ongoing effort. Indeed,
despite the Sloan Digital Sky Survey permitting the discovery of a
plethora of new ZZ Ceti stars, there are still relatively bright (V
$<$ 17) ZZ Ceti stars to be discovered as we and others
\citep{silvotti05,voss06} are finding out.

\begin{landscape}
\begin{figure}
\centering
\subfigure[]{\includegraphics[bb=50 35 600 700,height=9.5cm]{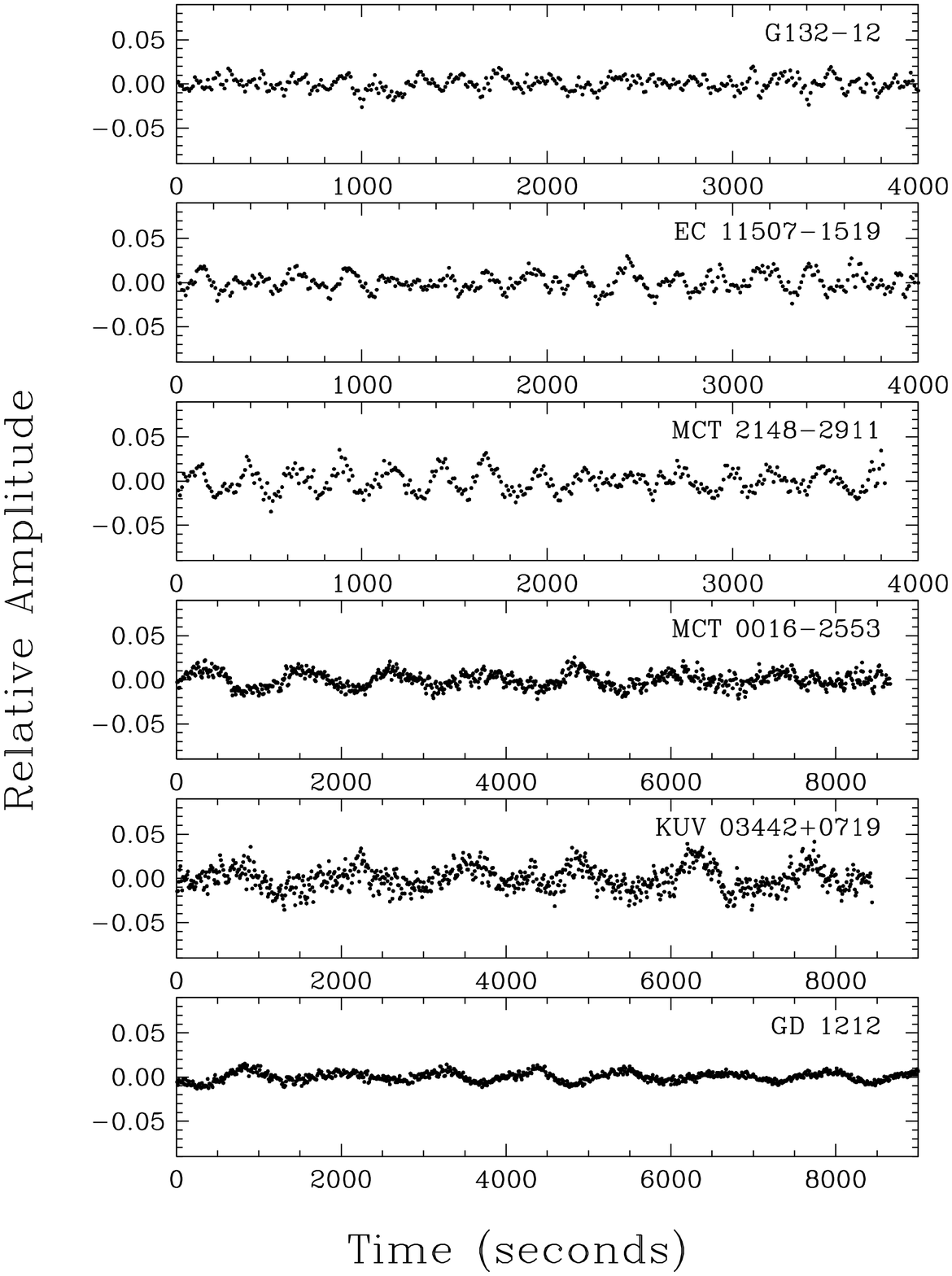}\label{lcurves}}
\subfigure[]{\includegraphics[bb=0 125 575 600,height=8.5cm]{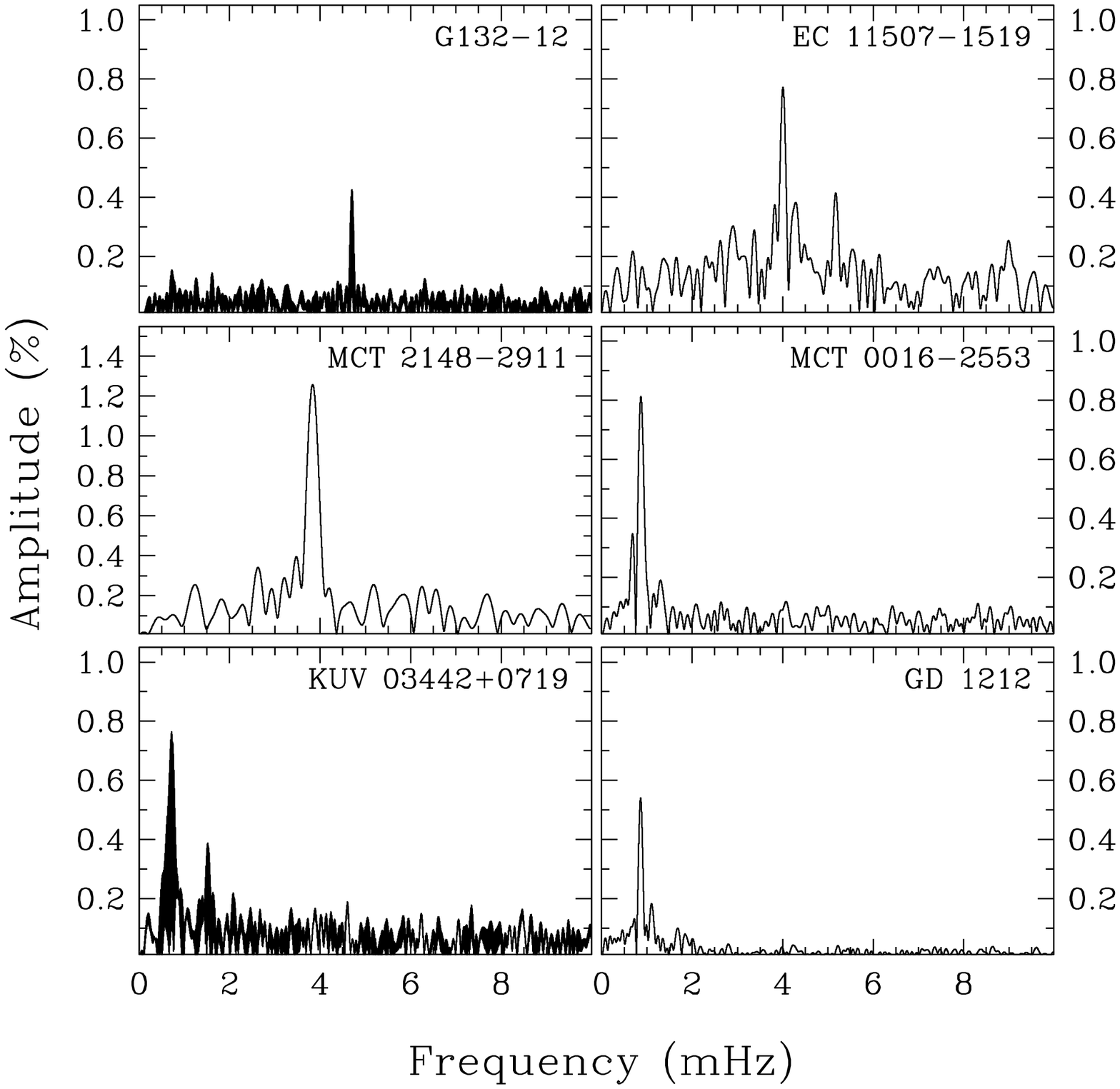}\label{ffts}}
\caption{(a) Light curves of the 6 new ZZ Ceti stars listed in Table
\ref{tab1}. Each point represents a sampling time of 10 s. The light
curves are expressed in terms of residual amplitude relative to the
mean brightness of the star. (b) Fourier (amplitude) spectra for the
light curves of the 6 new ZZ Ceti stars in the 0-10 mHz bandpass. The
spectrum in the region from 10 mHz to the Nyquist frequency is
entirely consistent with noise and is not shown. The amplitude axis is
expressed in terms of the percentage variations about the mean
brightness of the star.}
\end{figure}
\end{landscape}

\section{Photometric Observations \& Analysis}

\begin{table}[!t]
\caption{Photometric parameters of the six new ZZ Ceti white dwarfs}\label{tab1}
\begin{center}
\begin{tabular}{clccc}
\hline
\hline
   &       &         & Amplitude & Dominant   \\
WD &  Name & {\it V} &   (\%)    & Period (s) \\
\hline
\multicolumn{5}{c}{Short Period}\\
\hline
0036+312   & G132-12         & 16.20 & 0.43 &  212.7 \\
1150$-$153 & EC 11507$-$1519 & 16.00 & 0.77 &  249.6 \\
2148$-$291 & MCT 2148$-$2911 & 16.10 & 1.26 &  260.8 \\
\hline
\multicolumn{5}{c}{Long Period}\\
\hline
0016$-$258 & MCT 0016$-$2553 & 16.10 & 0.81 & 1152.4 \\
0344+073   & KUV 03442+0719  & 16.10 & 0.76 & 1384.9 \\
2336$-$079 & GD 1212         & 13.26 & 0.54 & 1160.7 \\
\hline
\end{tabular}
\end{center}
\end{table}

We proceeded to secure high-speed photometric measurements for those
stars in our spectroscopic survey whose atmospheric parameters placed
them inside or near the edges of the known ZZ Ceti instability
strip. In all, 6 of these stars turned out to be genuine pulsating DA
white dwarfs: ZZ Ceti stars. MCT 2148$-$2911 was observed through a
B-band filter on 2005 August 13 with the 3.6 m Canada-France-Hawaii
telescope equipped with LAPOUNE, the portable Montr\'eal three-channel
photometer. G132-12, MCT 0016$-$2553, KUV~03442+0719, and GD 1212 were
observed in ``white'' light during a 5 night run from 2005 October 23
to 27 at the Steward Observatory 2.3 m telescope, equipped once again
with LAPOUNE. Finally, EC 11507$-$1519 (see also Koester \& Voss,
these proceedings) was observed on 2006 March 22 at Steward
Observatory with the same telescope and setup. We also observed
another candidate star during this last run, WD 1149+057 (PG
1149+058). Unfortunately, due to increasing cloud cover, our light
curve for this object was inconclusive. It was later reported as a new
ZZ Ceti star by \citet{voss06}. Figure \ref{lcurves} displays the
sky-subtracted, extinction-corrected light curves obtained for each
star. The resulting Fourier (amplitude) spectra are shown in Figure
\ref{ffts}.

Summarized in Table \ref{tab1} are the data for the 6 new ZZ Ceti
stars split into short and long period variables. We list the $V$
magnitude as well as the amplitude and period of the dominant
pulsation mode obtained from the corresponding Fourier spectra in
Figure \ref{ffts}. We note that the periods of the dominant
oscillation modes in the first three stars of Table \ref{tab1}, as
well as those of the last three objects, are consistent with the
positions of these stars inside but near the blue (short-period
pulsators) and red (long-period pulsators) edges of the instability
strip, respectively. However, as a curiosity, we also point out that
the amplitudes of the detected modes in the three ``red edge''
pulsators are not very large, especially for GD 1212. This goes
against the general tendency to observe large amplitudes in the cooler
objects. Are we observing stars at the very red edge of the strip,
basically ``dying off'' in amplitude? Note also the very long period
seen in KUV 03442+0719, one of the longest if not the longest one ever
detected in a ZZ Ceti star \citep{mukadam06}.

To obtain proper time-averaged spectra for ZZ Ceti stars, it is
necessary to set the exposure time long enough to cover several
pulsation cycles. However, the exposure times for the optical spectra
of MCT 0016$-$2553, KUV 03442+0719 and GD 1212 are 1800, 1200 and 600
s respectively. By comparing with the dominant periods listed in Table
\ref{tab1}, we see that this criterion was not met for these three
long-period variables since we did not know a priori that these white
dwarfs would turn out to be variable. It will therefore be necessary
to acquire new optical spectra for these stars in order to refine our
determination of their atmospheric parameters. Consequently, the
atmospheric parameters for these objects are considered preliminary
\citep{gianninas06}.

\begin{table}[!b]
\caption{New and revised atmospheric parameters of ZZ Ceti and photometrically constant DA white dwarfs}\label{tab2}
\begin{center}
\begin{tabular}{clcccc}
\hline
\hline
     & $T_{\rm eff}$ &          &       M       &         &      \\
Name &      (K)      & $\log g$ & ($M_{\odot}$) & $M_{V}$ & Note \\
\hline
\multicolumn{6}{c}{New ZZ Ceti}\\
\hline
HS 1249+0426 & 12040 & 8.15 & 0.70 & 11.83 & 1 \\ 
HS 1531+7436 & 12920 & 8.45 & 0.90 & 12.18 & 1 \\
HS 1625+1231 & 11730 & 8.15 & 0.70 & 11.89 & 1 \\
HS 1824+6000 & 11380 & 7.82 & 0.51 & 11.50 & 1 \\
\hline
\multicolumn{6}{c}{New Photometrically Constant}\\
\hline
HS 1253+1033 & 13040 & 7.85 & 0.53 & 11.27 & 2 \\
HS 1544+3800 & 13550 & 7.95 & 0.58 & 11.34 & 1 \\
HS 1556+1634 & 11972 & 7.44 & 0.35 & 10.89 & 1 \\
\hline
\multicolumn{6}{c}{Revised ZZ Ceti}\\
\hline
WD 1149+057   & 11210 & 8.19 & 0.72 & 12.07 & 1,3 \\
WD 1349+552   & 12010 & 7.93 & 0.57 & 11.53 & 4   \\
WD 1429$-$037 & 11370 & 8.08 & 0.67 & 11.87 & 2,4 \\
WD 1541+650   & 11640 & 8.18 & 0.72 & 11.95 & 4   \\
\hline
\end{tabular}
\end{center}
Notes. --
(1) Photometric status reported in \citet{voss06}
(2) Photometric status reported in \citet{silvotti05}
(3) Time-averaged spectrum
(4) Improvement of S/N (S/N $>$ 70)
\end{table}

\begin{center}
\begin{figure}[!ht]
\includegraphics[bb=75 50 575 700,height=12cm,angle=270]{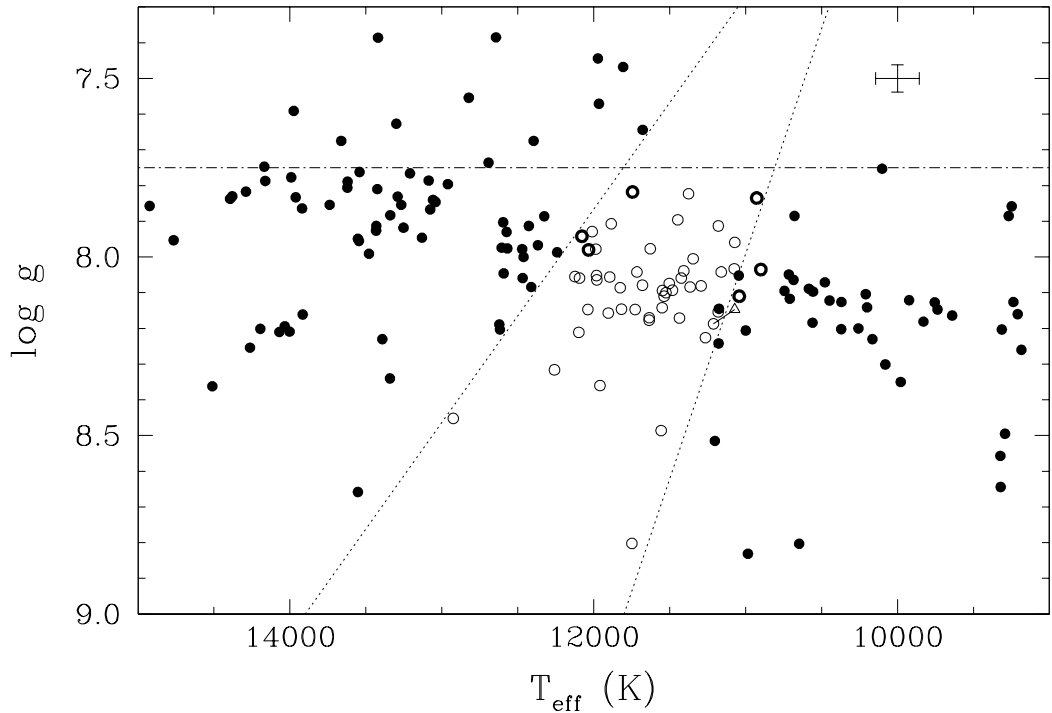}
\caption{{Region of the $T_{\rm eff}$-$\log g$ plane containing the ZZ
Ceti instability strip. The error bars in the upper right corner
represent the average uncertainties of the spectroscopic method in the
region of the ZZ Ceti stars (1.2\% in $T_{\rm eff}$ and 0.038 dex in
$\log g$)}}\label{strip}
\end{figure}
\end{center}

\section{Latest Spectroscopic Results}

During an observing run in 2006 April at the Observatoire du mont
M\'egantic, we were able to secure high signal-to-noise ratio (S/N)
optical spectra for a number of the stars reported by
\citet{silvotti05} and \citet{voss06} as either variable or
non-variable. We also obtained higher S/N ratio spectra for several ZZ
Ceti stars in our sample as well as a properly time-averaged spectrum
for the ZZ Ceti star WD 1149+057. The atmospheric parameters we
measured for these stars are listed in Table \ref{tab2} along with the
stellar masses and absolute visual magnitudes. Our theoretical
framework and fitting technique are described at length in
\citet{gianninas05}. In this case, we knew beforehand the photometric
nature of our targets and thus we set our exposure times long enough
to cover at least 3 pulsation cycles for each star.

We can now update our previous photometric sample \citep{gianninas06}
by appending these 4 new ZZ Ceti white dwarfs and 3 photometrically
constant DA stars as well as incorporating the revised values of
$T_{\rm eff}$-$\log g$ for the last 4 objects listed in Table
\ref{tab2}. Thus, our current view of the empirical ZZ Ceti
instability strip is shown in Figure \ref{strip}. The black open and
filled circles represent ZZ Ceti stars and photometrically constant DA
stars respectively. The bold open circles are the 6 new ZZ Ceti stars
from Table 1. The open triangle is the old position of WD 1149+057 in
the $T_{\rm eff}$-$\log g$ plane, it is connected to its new location
by a straight line. The dashed lines represent our empirical
determination of the blue and red edges of the instability strip.

It is well known that white dwarfs in this temperature range that are
less massive than $\sim$ 0.475 $M_{\odot}$ ($\log g \sim$ 7.75) must
necessarily be the product of binary evolution. This limit is
represented by the dashed-dotted line in Figure \ref{strip}. Indeed,
several of the stars which lie above this line have been discovered as
being part of unresolved double degenerate systems
\citep{maxted99,maxted00}. In such cases, the atmospheric parameters
obtained are an average of the parameters of both components of the
system. Thus, the atmospheric parameters for stars that lie above the
dashed-dotted line are considered very uncertain and they cannot be
used to constrain the boundaries of the instability strip. Conversely,
HS 1531+7436, now the hottest ZZ Ceti star in our sample, allows us to
pin down the location and slope of the blue edge of the ZZ Ceti
instability strip. However, more hot ZZ Ceti stars lying near the
empirical blue edge are needed in order to firmly establish its
location. Furthermore, since the atmospheric parameters for the new
long period variables are preliminary, we have refrained from
redefining the empirical red edge. We also notice that the atmospheric
parameters derived from a properly time-averaged spectrum of WD
1149+057 place it within the confines of our empirical boundaries. We
expect the same result once we have secured new spectra for the new
long-period ZZ Ceti stars. Finally, we see that the instability strip
remains pure, devoid of any photometrically constant stars within its
empirical boundaries.

\section{Conclusions \& Future Work}

We have discovered 6 new and relatively bright ZZ Ceti stars and have
been able to constrain rather well the empirical boundaries of the ZZ
Ceti instability strip. This brings us one step closer to an
understanding of the ZZ Ceti phenomenon as a whole but there is still
work to be done. Our spectroscopic survey of \citet{mccook99} is yet
to be completed, with $\sim$ 100 stars for which we still need to
acquire spectra. We also need to obtain time-averaged spectra for the
long-period variables we discovered. In addition, we want to continue
improving the precision of our $T_{\rm eff}$ and $\log g$
determinations by ensuring that all our spectra have S/N $>$
70. Finally, with the blue and red edges well constrained, we wish to
revisit pulsation theory as well as evolutionary models for DA white
dwarfs in order to produce the best match between our empirical result
and theoretical predictions.

\acknowledgements 
We would like to thank the director and staff of Steward
Observatory and the Canada-France-Hawaii Telescope for the use of
their facilities and for supporting LAPOUNE as a visitor
instrument. We would also like to acknowledge the financial support of
the Royal Astronomical Society. This work was supported in part by the
NSERC Canada. A. G. acknowledges the contribution of the Canadian
Graduate Scholarships. P. B. is a Cottrell Scholar of Research
Corporation. G. F. acknowledges the contribution of the Canada
Research Chair Program.


\end{document}